\documentclass{elsart41}
\usepackage{graphics}
\usepackage{graphicx}
\usepackage{amssymb}

\def\e{\varepsilon}
\def\as{\alpha\sigma}
\def\ket#1{|#1\rangle}
\def\bra#1{\langle #1|}

\begin{document}

\begin{frontmatter}



\title{Unusual Non-Fermi Liquid Behavior of
Ce$_{1-x}$La$_{x}$Ni$_{9}$Ge$_4$ Analyzed  in a Single Impurity
Anderson Model with Crystal Field Effects}


\author[CPM]{E.--W. Scheidt\corauthref{Scheidt}},
\ead{Scheidt@physik.uni-augsburg.de}
\author[CPM]{F. Mayr},
\author[CPM]{U. Killer},
\author[CPM]{W. Scherer},
\author[Au]{H. Michor},
\author[Au]{E. Bauer},
\author[Ge2]{S.~Kehrein},
\author[Goe]{Th. Pruschke},
\author[Bre]{F. Anders},

\address[CPM]{CPM, Inst. f. Physik,
Universit\"{a}t Augsburg, 86159 Augsburg, Germany}
\address[Au]{Inst. f. Festk\"orperphysik, TU Wien, 1040 Wien,
Austria}
\address[Ge2]{TP III -- EKM,
Inst. f. Physik, Universit\"at Augsburg, 86135 Augsburg, Germany}
\address[Goe]{Inst. f\"ur Theo. Physik, Universit\"at
G\"ottingen, 37077 G\"ottingen, Germany}
\address[Bre]{Inst. f. Theo. Physik,
Universit\"at Bremen, 28334 Bremen, Germany}

\corauth[Scheidt]{E.--W. Scheidt. Tel: +49\,821\,5983356 fax:
+49\,821\,5983227}

\begin{abstract}

CeNi$_{9}$Ge$_4$ exhibits unusual non-Fermi liquid behavior with
the largest ever recorded value of the electronic specific heat
$\Delta C/T \cong 5.5$\,JK$^{-2}$mol$^{-1}$ without showing any
evidence of magnetic order. Specific heat measurements show that
the logarithmic increase of the Sommerfeld coefficient flattens
off below 200\,mK. In marked contrast, the local susceptibility
$\Delta\chi$ levels off well above 200\,mK and already becomes
constant below 1\,K. Furthermore, the entropy reaches 2$R$ln2
below 20\,K corresponding to a four level system. An analysis of
$C$ and $\chi$ was performed in terms of an $SU(N=4)$ single
impurity Anderson model with additional crystal electric field
(CEF) splitting. Numerical renormalization group calculations
point to a possible consistent description of the different low
temperature scales in $\Delta c$ and $\Delta \chi$ stemming from
the interplay of Kondo effect and crystal field splitting.

\end{abstract}

\end{frontmatter}


Strongly correlated electron materials have been of interest for
many years since they allow to test the limits of Landau's Fermi
liquid (FL) theory. Specific heat and susceptibility measurements
are good tools  to distinguish whether electronic correlations
renormalize the Fermi liquid parameters or lead to a new non-Fermi
liquid (nFL) state. The breakdown of FL~theory and the borderline
between these two regimes continues to attract much interest
\cite{stewart01}. Recently, we have shown that
Ce${}_{1-x}$La${}_x$Ni${}_9$Ge${}_4$  is a very interesting system
for studying this borderline \cite{Killer04,Scheidt05}.

\begin{figure}[!ht]
\begin{center}
\includegraphics[width=0.35\textwidth]{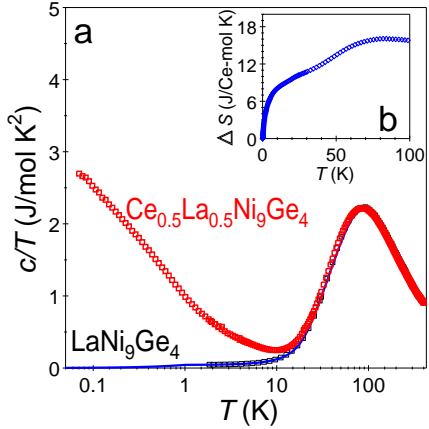}
\end{center}
\caption{(a)  A semi-logarithmic plot of $c/T$\,\emph{vs.}\,$T$ of
Ce$_{0.5}$La$_{0.5}$Ni${}_9$Ge${}_4$ and LaNi${}_9$Ge${}_4$. The
solid line mirrors the phonon contribution of LaNi${}_9$Ge${}_4$
(see text). (b) The electronic part of the entropy of
Ce$_{0.5}$La$_{0.5}$Ni${}_9$Ge${}_4$ as calculated from the data
of Fig.~\ref{fig2}\,a.} \label{fig1}
\end{figure}

The stoichiometric system ($x=0$) shows the largest ever recorded
value of the electronic specific heat $\Delta C/T \cong
5.5$\,JK$^{-2}$mol$^{-1}$ without any magnetic order
\cite{michor04}. This behavior is mainly driven by single ion (Ce)
effects \cite{Killer04}: the low-temperature behavior of the
Sommerfeld coefficient ($\gamma \sim \Delta c$($T$)$/T$) and the
local susceptibility ($\chi(T)$) are proportional to $1-x$ (except
for $x=0$ which shows additional collective effects). As
representative of all dilute samples we discuss
Ce$_{0.5}$La$_{0.5}$Ni$_{9}$Ge$_4$ in this paper. In order to
extract the electronic contribution to its specific heat, we
calculated the phonon contribution of the non \emph{f}-electron
system LaNi${}_9$Ge${}_4$. The solid line in Fig.~\ref{fig1}a is a
phonon--fit which is parametrized using a Debye term
($\Theta_{\rm{D}} = 135$\,K, 3 degrees of freedom (dof)) and two
Einstein modes ($\Theta_{\rm{E}} = 154$\,K, 12 dof;
$\Theta_{\rm{E}} = 301$\,K, 27 dof). This result is in good
agreement with recent neutron scattering measurements, where the
inelastic response in both CeNi${}_9$Ge${}_4$ and
LaNi${}_9$Ge${}_4$ was dominated by strong phonon peaks centered
at 15 and  35\,meV \cite{michor05}.

\begin{figure}[t]
\begin{center}
\includegraphics[width=0.45\textwidth]{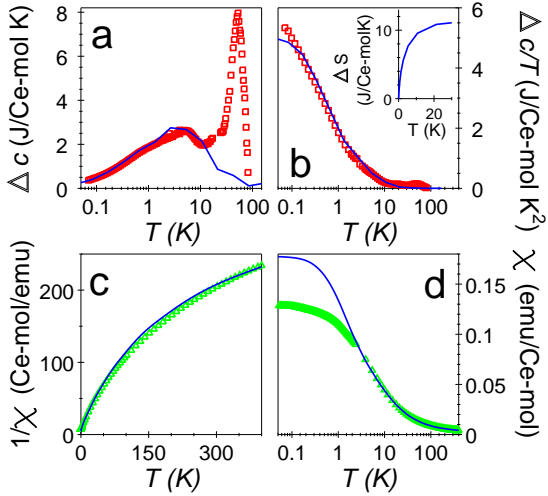}
\end{center}
\caption{ Ce$_{0.5}$La$_{0.5}$Ni${}_9$Ge${}_4$: (a) a
semi-logarithmic plot of the electronic contribution of the
specific heat $\Delta c$ and (b) $\Delta c/T$. (c) The inverse
susceptibility \emph{vs.} $T$ and (d) the local susceptibility
down to 30\,mK.  The lines always show the fit results from the
NRG calculation (see text).} \label{fig2}
\end{figure}

Notice that the Sommerfeld coefficient $\Delta c/T$ of
Ce${}_{0.5}$La${}_{0.5}$Ni${}_9$Ge${}_4$ shows nFl-behavior
(logarithmic increase) down to 60\,mK, while $\chi$($T$) becomes
constant (Fermi liquid like) below 1\,K (Fig.~\ref{fig2}). From an
entropy calculation ($S=2R\,\rm{ln}2$ for $T < 20$\,K) one could
suggest that the crystal electrical field (CEF) ground state
quartet of Ce$^{3+}$ splits into two doublets leading to an
interplay between Kondo~effect and CEF--splitting on the same
energy scale \cite{Killer04}. In this work we use a numerical
renormalization group calculation (NRG) to establish a consistent
description of the different low temperature scales in $\Delta c$
and $\Delta \chi$. We use the SU(4)--Anderson impurity model with
additional crystal field splitting as a possible theoretical
minimal model,
\begin{eqnarray}
H&=& H_{cond} + H_{imp} + H_{hyb} \\
  H_{cond}&=& \sum_{\sigma}\sum_{\alpha=1,2} \e_{k\as} c^\dagger_{k\as}
  c_{k\as}
\\
H_{imp} &=& \sum_{\as} E_{\alpha} \ket{\as}\bra{\as}
\\
H_{hyp} &=& \sum_{k\as} V_{\as} \left(c^\dagger_{k\as}
  \ket{0}\bra{\as}
+ \ket{\as}\bra{0} c_{k\as}
\right) \ .
\end{eqnarray}
$H_{imp}$ describes the dynamics of the Ce 4\emph{f}-states under
the assumption that only the unoccupied state $\ket{0}$ and singly
occupied states $\ket{\as}$ play a role for the low--temperature
physics. Therefore the Schottky peak at about 60\,K due to the
$J=5/2$ states is necessarily absent in the NRG fits (see
Fig.~\ref{fig2}a). We found the best agreement with the
low--temperature data for a crystal field splitting
$\Delta=E_2-E_1\approx 10$\,K for the ground state doublet
$\ket{\as}$ and approximately the same Kondo temperature $T_{K}$
due to the hybridization matrix elements $V_{\as}$ \cite{tbp}.
From Fig.~\ref{fig2}c and~2d one concludes that while the
quantitative agreement at very low temperatures is not yet
satisfactory, the interplay of crystal field effects and Kondo
physics can lead to temperature scales for the onset of Fermi
liquid like behavior in the specific heat and magnetic
susceptibility that differ by a factor of about~10. This would
imply that such an interplay can result in an extended NFL--like
logarithmic behavior of $\Delta c/T$, while eventually the system
does become a Fermi liquid at even lower temperatures.

This observation based on our NRG calculations is of general importance
for analyzing NFL--like behavior in heavy fermion
systems. Whether our theoretical model can quantitatively explain the experimental data
for Ce${}_{1-x}$La${}_x$Ni${}_9$Ge${}_4$,
however, still
needs to be explored further. Work along these lines is in progress.

%
%
%
%

This work was supported by the SFB~484 of the Deutsche
Forschungsgemeinschaft (DFG).


\begin{thebibliography}{99}
\bibitem{tbp}
Details of this calculation will be published separately.
\end{thebibliography}
\end{document}